\documentclass[12pt]{article}
\usepackage{appendix}
\usepackage{graphicx}
\usepackage{amsmath}
\usepackage{amsfonts}
\usepackage{color}    
\usepackage{bm}
\usepackage{subfigure}
\usepackage{epstopdf}
\usepackage{authblk}

\begin{document}

\title{Exploiting tumor shrinkage through temporal optimization of radiotherapy}

\author{Jan Unkelbach$^1$, David Craft$^1$, Theodore Hong$^1$, \\
D{\'a}vid Papp$^1$, Jagdish Ramakrishnan$^2$, Ehsan Salari$^3$, \\
John Wolfgang$^1$, Thomas Bortfeld$^1$}
\affil{$^1$ Department of Radiation Oncology, Massachusetts General Hospital and Harvard Medical School, Boston, MA, USA}
\affil{$^2$ Wisconsin Institute for Discovery, University of Wisconsin, \\ Madison, WI, USA}
\affil{$^3$ Department of Industrial and Manufacturing Engineering, \\ Wichita State University, Wichita, KS, USA}


\maketitle

\begin{abstract}

Purpose: In multi-stage radiotherapy, a patient is treated in several stages separated by weeks or months. This regimen has been motivated mostly by radiobiological considerations, but also provides an approach to reduce normal tissue dose by exploiting tumor shrinkage. The paper considers the optimal design of multi-stage treatments, motivated by the clinical management of large liver tumors for which normal liver dose constraints prohibit the administration of an ablative radiation dose in a single treatment.

Method: We introduce a dynamic tumor model that incorporates three factors: radiation induced cell kill, tumor shrinkage, and tumor cell repopulation. The design of multi-stage radiotherapy is formulated as a mathematical optimization problem in which the total dose to the normal tissue is minimized, subject to delivering the prescribed dose to the tumor. Based on the model, we gain insight into the optimal administration of radiation over time, i.e. the optimal treatment gaps and dose levels. 

Results: We analyze treatments consisting of two stages in detail. The analysis confirms the intuition that the second stage should be delivered just before the tumor size reaches a minimum and repopulation overcompensates shrinking. Furthermore, it was found that, for a large range of model parameters, approximately one third of the dose should be delivered in the first stage. The projected benefit of multi-stage treatments in terms of normal tissue sparing depends on model assumptions. However, the model predicts large dose reductions by more than a factor of two for plausible model parameters. 

Conclusions: The analysis of the tumor model suggests that substantial reduction in normal tissue dose can be achieved by exploiting tumor shrinkage via an optimal design of multi-stage treatments. This suggests taking a fresh look at multi-stage radiotherapy for selected disease sites where substantial tumor regression translates into reduced target volumes.

\end{abstract}

\section{Introduction}

Radiotherapy planning aims at maximizing the chance of cancer cure while minimizing the risk of side effects in normal tissues. This goal is approached by optimizing the dose distribution in both space and time. The spatial aspect of radiotherapy planning amounts to adequate target volume delineation and subsequent optimization of the spatial dose distribution, e.g. through intensity-modulated radiotherapy (IMRT). The temporal aspect of treatment planning considers the question how radiotherapy is administered over time.\\ 

In this work, we consider a specific application of temporal optimization in radiotherapy. The approach aims at reducing normal tissue dose by exploiting tumor shrinkage over the course of therapy. The motivational example comes from a patient who presented with a very large liver tumor. It was not possible to deliver an ablative radiation dose without exceeding commonly used dose-volume constraints for the non-involved liver. The patient was treated at our institution in two stages, with a four months gap in between. By waiting for 4 months after the first stage, we could take advantage of significant tumor shrinkage during the gap period, in this case down to less then 50\% of the original volume. This allowed us to shrink the treatment fields at the second stage of treatment, thereby reducing the dose burden on surrounding healthy organs. A general concern with this unconventional treatment regimen is of course the growth of the residual tumor during the treatment gap period. The purpose of this paper is to understand the potential advantage of this approach by introducing a simple biological model that takes the tradeoff between tumor shrinkage and tumor cell repopulation into account. We also wish to maximize the advantage by optimizing the time gap and the dose delivered at each treatment stage. \\

The biological motivation for exploiting tumor shrinkage in multi-stage treatments stems from the fact that the standard radiation treatment regime, in which a fixed dose of radiation is delivered consecutively and in the same way over a number (5-40) of treatment days, is inefficient. Even after only one treatment fraction of 2 Gy or more, the number of active tumor cells is typically already reduced by about a factor of 2. In all subsequent fractions, we therefore treat mostly tumor cells that are already dead. Near the end of the treatment course, only a tiny number of active tumor cells remains. Tumor shrinkage, which may occur because macrophages clear out the dead tumor cells over time, provides an opportunity to avoid treating the dead cells and thereby increase the efficiency of the use of radiation substantially. \\

While we have found one paper that shrinks the field size due to tumor regression \cite{hayakawa1992}, it seems that the primary motivation for the rest period is to allow the normal tissue to recover, rather than waiting for the tumor to regress further. The Nobel lecture by Hounsfield in 1979 \cite{hounsfield1980} describes in qualitative terms what we model in this paper.  Nowadays, modern technologies may allow us to clinically implement a multi-stage radiotherapy approach to exploit tumor shrinkage. Advances in imaging allow for a better delineation of the tumor. In addition, the precision in dose delivery has improved substantially with the development of intensity-modulated radiotherapy, proton therapy, and cone beam imaging for patient setup. 

\paragraph{Review of multi-stage radiotherapy:} The idea of dividing a treatment into two or more phases with a rest period of typically 2-4 weeks has been around for a long time. Such treatments have been called {\it split-course} radiotherapy and have been discussed as early as 1935 \cite{holsti1969}. The most common argument for a split-course is better preservation of normal tissue due to its faster regeneration capability compared to tumors in the rest period. Others include better treatment efficacy due to improved tumor vascularization and increased tumor re-oxygenation after the rest period. The rationale has therefore been purely radiobiological and has possibly kept the rest period short and in the order of 2-4 weeks. The efficacy of split-course therapy has been assessed by many randomized trials and retrospective studies \cite{dubben2001}. There have been studies using split-course for tumors of the head and neck \cite{holsti1969}, lung \cite{holsti1980,hayakawa1992}, and pelvis \cite{marcial1985}. However, to the best of our knowledge, there do not seem to have been any for liver tumors. Several randomized studies in the 1960s through 1980s had indicated split-course radiotherapy to be no worse than conventional continuous treatment \cite{dubben2001}. However, later retrospective studies have indicated that split-course treatments are less effective, especially for fast growing tumors such as head and neck \cite{akimoto1997,overgaard1988}. Further discussion about previous split-course radiotherapy studies can be found in \cite{dubben2001}. These previous works, however, have not used a smaller field size at the second stage, meaning there was no physical dose reduction for the normal tissue.

\paragraph{Previous work on temporal optimization:} Previous work on optimizing radiation delivery over time has been performed in the context of fractionation. The radiobiological basis of fractionation are repair processes, which happen within hours after irradiation. Thus, work on the optimization of fractionation schemes is mostly based on the linear-quadratic cell survival model and extensions thereoff. It is known that time dependencies such as incomplete repair in between fractions and accelerated repopulation give rise to nontrivial nonuniform fractionation schemes \cite{brenner1994optimizing,wein2000dynamic,yang2005optimization,bertuzzi2013optimal,bortfeld2013optimization}. Even though our paper is also concerned with temporal optimization of radiotherapy, it is based on tumor shrinkage which happens on a time scale of several months. Instead, fractionation is based on repair processes that happen on a time scale of hours. Thus, these two applications are based on distinct processes and should be distinguished.

\paragraph{Contribution and organization of this paper:} The paper has two specific objectives: First, we want to estimate the amount of normal tissue sparing that can be expected from exploiting tumor shrinkage in a multi-stage treatment. Second, we want to gain insight into the optimal treatment regimen, i.e. the optimal doses delivered in each stage and the optimal time gaps between stages. The rest of this paper is organized as follows: Section \ref{SecCase} introduces the clinical case and treatment strategy that served as the motivation for this study. In section \ref{SecModel} we develop the model of tumor growth and response to radiation. We further formulate the design of multi-stage treatments as a mathematical optimization problem. Sections \ref{Sec:2stages} and \ref{SecSensitivity} analyze treatments with two stages in detail. In section \ref{Sec:nstages}  we discuss the benefit of adding a third stage, and obtain qualitative insight into the optimal $N$-stage treatment. Finally, section \ref{SecDisc} summarizes the results and discusses the clinical application of multi-stage radiotherapy, focusing on liver metastases.



\section{Clinical motivation: large liver tumors}
\label{SecCase}
The most commonly used dosing strategy for liver stereotactic body radiation therapy (SBRT) is based on dose-volume constraints for the non-involved liver. Thereby, the prescription dose depends on the amount of radiation the non-involved liver receives, and is limited such that the normal liver constraints are fulfilled.  With this paradigm, as the ratio of tumor to non-involved liver increases, the amount of radiation that can be given decreases. This leads to a paradox: larger tumors are treated with lower doses.  Thus, in patients with very large tumors, SBRT with the current dosing paradigms is unlikely to be ablative and cannot offer potentially curative therapy. Thus, from a clinician’s perspective, a patient with a large solitary liver lesion cannot be cured, in spite of the absence of extrahepatic disease. Therefore, there is a clinical motivation to find alternative strategies for dealing with large lesion cases.

\subsection{Case report}
\begin{figure*}[htb]
\centering
\subfigure[Stage 1]{
\includegraphics[height=5.5cm]{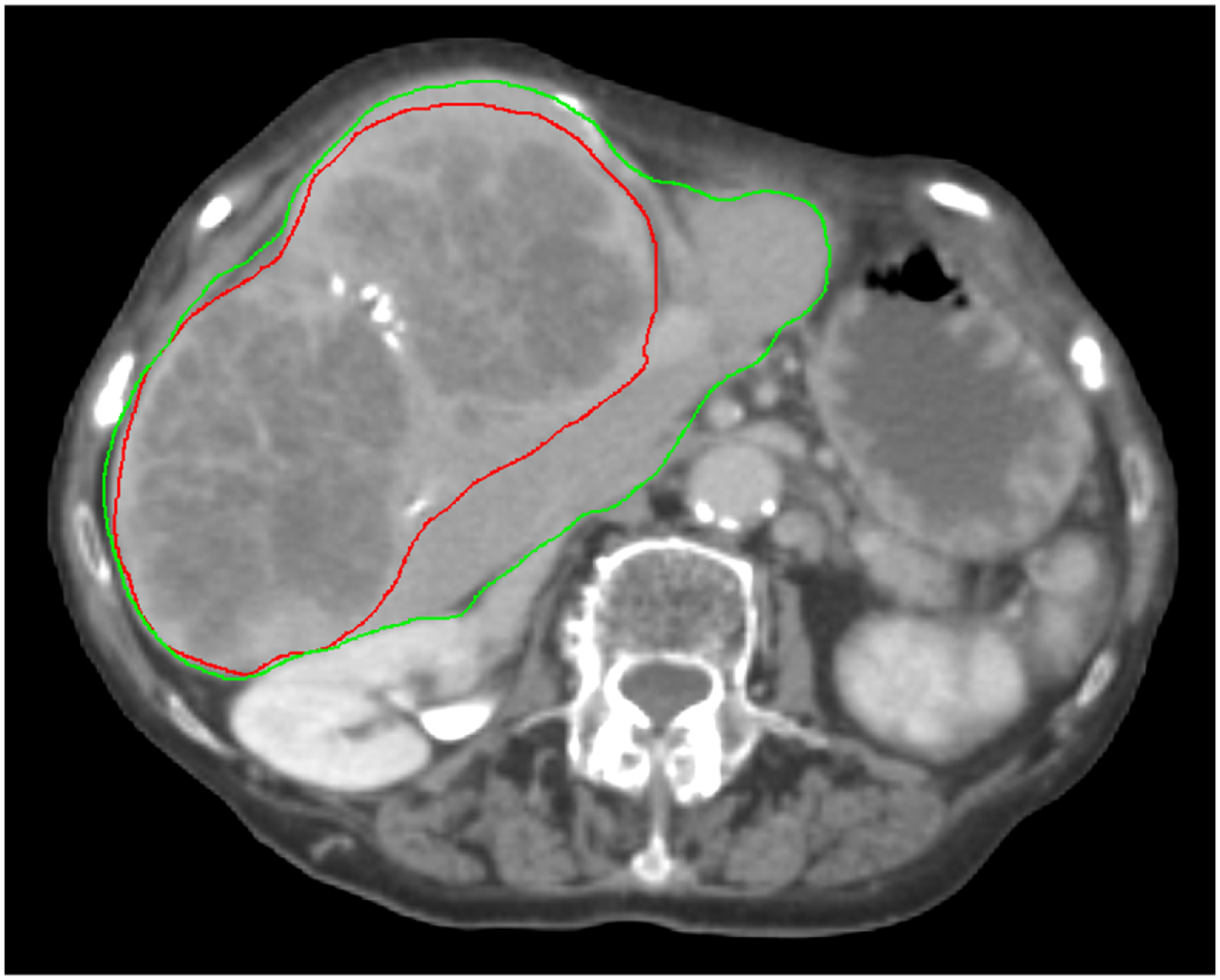}
}
\subfigure[Stage 2]{
\includegraphics[height=5.5cm]{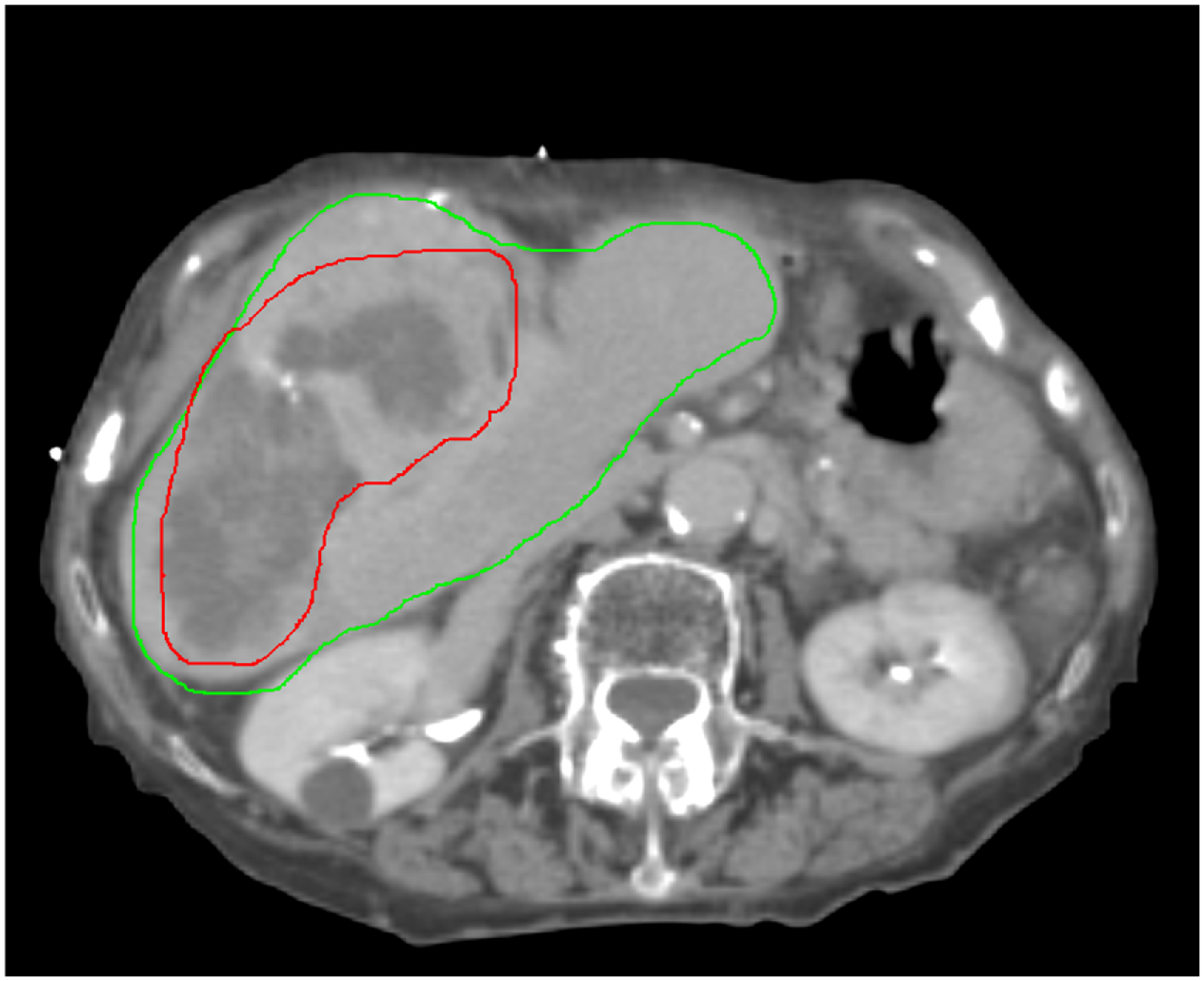}
}
\caption{Planning CT image of the patient for the first treatment stage (a) and the second stage (b) four months later. The corresponding CT slices for the two time points are chosen based on the bony anatomy of the spine. The red contour shows the GTV, the green contour corresponds to the liver.}
\label{Fig:CT}
\end{figure*}

The patient who motivated this work presented with a very large liver tumor (a metastatic lesion originating from chemo-refractory colorectal cancer). The initial target volume in the first treatment stage was assessed using 4D CT, resulting in an internal gross target volume (IGTV) of 1218 cc in size (which incorporated a peak-to-peak motion of 4.15 mm as defined at the fiducials). In comparison, the total volume of normal appearing liver was 870 cc. Thus, the tumor mass accounted for 58\% of the total liver volume. The patient was treated in two stages separated temporally by a period of four months. Figure \ref{Fig:CT} shows the radiotherapy planning CT for the first stage (a) and the second stage (b). Until the most recent follow-up visit (6 months after the second stage), the liver lesion remains stable. \\

Both treatment stages used a dose painting strategy in which the planning target volume was subdivided into two parts: a low dose PTV was constructed via a 3 mm isotropic expansion of the IGTV; a high dose PTV was constructed by a 1 cm contraction of the low dose PTV. This allowed for a dose reduction to normal radiosensitive tissue adjacent to the target volume while delivering a high SBRT dose to the interior target volume. In the first stage the low dose PTV was treated to 18 Gy in 3 fractions while the high dose PTV was treated to a total of 36 Gy in 3 fractions concurrently.  After the four month intermission, the patient returned to be rescanned and treated using the new imaging information. The physician-defined IGTV was 499 cc (37\% of total liver volume), and thereby substantially smaller than in the first stage (see figure \ref{Fig:CT}b). The prescribed dose to the low dose PTV for stage two was 20 Gy in 3 fractions, while the high dose PTV received 30 Gy in 3 fractions. Deformable registration was performed in order to assess the cumulative dose to secondary organs at risk, such as the spinal cord and the kidneys. Through the dose painting approach, the mean dose to the non-involved liver (i.e. total liver without GTV) was reduced to 10.8 Gy in the first stage, and 13.3 Gy in the second stage.

\section{Model for staged tumor irradiation}
\label{SecModel}

\subsection{Model of tumor growth and radiation response}
\label{SecTumorModel}
In this section, we formulate a model that describes the temporal evolution of the tumor size due to growth and radiation response. We assume that the size of the tumor is influenced by three factors:
\begin{itemize}
\item[1.] Radiation therapy: The radiation effect on the tumor is described by exponential cell kill such that the surviving fraction of tumor cells after irradiation is given by  $ e^{-\alpha d}$. Here, $d$ is the total physical dose delivered, and $\alpha$ is an effective radiosensitivity parameter which incorporates the fractionation scheme to be used\footnote{Assuming that the dose $d$ is delivered using $n$ fractions and a dose per fraction $d_f$, the linear-quadratic cell survival model can be written as $S=\exp ( - n (\alpha_0 d_f + \beta_0 d_f^2 )) = \exp (-\alpha d)$ where $\alpha = \alpha_0 (1 + d_f / (\alpha_0 / \beta_0))$ is an effective radiosensitivity parameter that depends on the $\alpha / \beta$-ratio and increases linearly with the dose per fraction. The envelope of the log cell survival curve thus becomes linear in the total dose.}. 
\item[2.] Repopulation: Surviving tumor cells after radiotherapy are assumed to proliferate exponentially. After a time $t$, the number of active tumor cells increases by a factor of $e^{t/\tau_g}$ where $\tau_g$ is the time constant for tumor repopulation.
\item[3.] Tumor shrinkage: Tumor cells that are lethally damaged during irradiation are slowly removed. It is assumed that this process is described by an exponential decay \cite{lim2008cervical,huang2010predicting}. After a time $t$ the fraction of dead tumor cells that remain is given by $e^{-t/\tau_s}$ where $\tau_s$ is the time constant for tumor shrinkage.
\end{itemize}
The entire treatment consists of $N$ treatment stages which we index by $i$. At each stage, a dose $d_i$ is delivered to the tumor. After completion of stage $i$, a treatment gap $t_i$ is imposed before the next stage is delivered. At the beginning of treatment stage $i$, the tumor consists of both viable tumor cells and dead tumor cells that have not been removed yet. We denote the number of viable tumor cells by $a_i$ and the number of dead tumor cells by $z_i$. The discrete time evolution equations for both cell compartments is governed by state update equations. Based on the above assumptions, the number of viable tumor cells at the next stage is given by 
\begin{equation}
a_{i+1} =  a_i  \, \underbrace{\exp(-\alpha d_i)}_{\mbox{cell kill}}
\, \underbrace{\exp(t_i /\tau_g)}_{\mbox{repopulation}}
\label{EqState_a}
\end{equation}
The change in the number of dead tumor cells is given by
\begin{equation}
z_{i+1} = \underbrace{z_i \exp (-\frac{t_i}{\tau_s} )}_{\mbox{further decay}} + \underbrace{a_i \left( 1-\exp(-\alpha d_i ) \right)\exp (-\frac{t_i}{\tau_s} )}_{\mbox{decay of cells killed in stage $i$}}
\label{EqState_z}
\end{equation}
where the first term describes the further decay of the dead tumor cells $z_i$ that were present at stage $i$, and the second term describes the decay of the tumor cells killed in stage $i$. The tumor volume $v_i$, i.e. the current tumor volume in stage $i$, is given by the sum of viable and dead tumor cells:
\begin{equation}
v_i = a_i + z_i 
\end{equation}
In this paper, we are primarily interested in relative changes. Therefore, we normalize $a_i$ and $z_i$ to the initial number of viable tumor cells. Consequently, $v_i$ represents the relative tumor volume in stage $i$ compared to the initial tumor volume. The initial state is thus given by
\begin{equation}
a_1=1 \quad z_1=0 \quad v_1=1
\end{equation}

\subsection{Normal tissue sparing factor}
For the initial treatment stage in which dose $d$ is delivered to the tumor, the normal tissue receives an inhomogeneous dose distribution with mean dose $\delta d$. We refer to $\delta < 1$ as the initial sparing factor for the normal tissue. For our work, we have to make assumptions on how the sparing factor decreases while the tumor shrinks. We assume that the sparing factor decreases proportionally with the area of the radiation field needed to treat the tumor. In turn, the area of the radiation fields decreases quadratically with the diameter of the tumor, while tumor volume decreases cubically. This leads to the following generic model for the sparing factor $\delta_i$ in stage $i$:
\begin{equation}
\delta_i = \delta \cdot v_i^{2/3}
\label{Eq:delta}
\end{equation}
This generic model has limitations since it does not take the specific patient geometry and the beam arrangement into account. Therefore, we use this generic model in the main text of the manuscript, but in section \ref{SecSensitivityQ} we discuss the sensitivity of our results with respect to deviations from this model.

\subsection{Optimization of multi-stage treatments}
In the optimal design of general $N$ stage treatments, we aim to determine the doses $d_i$ and the time gaps $t_i$ for each stage. The goal consists of minimizing the cumulative mean dose delivered to the healthy tissue, while assuring that a desired surviving fraction of viable tumor cells is achieved by the end of treatment\footnote{In this formulation, we assume a homogeneous dose prescribed to the tumor, and we consider the mean dose in the normal tissue. Considering the mean dose is appropriate for tumors embedded in a dose-limiting organ, such as liver tumors. In addition, these two quantities do not rely on dose accumulation on a voxel-by-voxel basis, which would be difficult to perform due to tumor shrinkage and large deformations of the normal tissue.}. This can be formulated as the following optimization problem:
\begin{eqnarray} 
\underset{d_i \ge 0, ~t_i \ge 0}{\mbox{minimize}} &~& \sum_{i=1}^{N} \delta_i d_i  \label{EqObj}\\
\mbox{subject to} &&  \nonumber 
\end{eqnarray}
\begin{eqnarray} 
\sum_{i=1}^N d_i &=& D^P + \underbrace{\frac{\sum_{i=1}^{N-1} t_i}{\alpha\tau_g}}_{\mbox{repopulation}}  \label{EqDoseCons} \\
a_{i+1} &=&  a_i  \exp(-\alpha d_i)\exp(\frac{t_i}{\tau_g}) \label{a} \\ 
z_{i+1} &=& z_i \exp(-\frac{t_i}{\tau_s}) + a_i \left( 1-\exp(-\alpha d_i ) \right)\exp(-\frac{t_i}{\tau_s})  \label{z} \\ 
v_i &=& a_i + z_i \\
\delta_i &=& \delta  v_i^{2/3} \label{delta}\\
a_1  &=&1,~~ z_1 = 0 \label{EqInitial}
\end{eqnarray}
The objective function (\ref{EqObj}) to be minimized represents the cumulative mean normal tissue dose over the duration of the treatment. The constraint in equation (\ref{EqDoseCons}) determines the total dose delivered to the tumor: $D^P$ denotes the tumor dose that would be prescribed in a single stage treatment; the second term represents the extra dose needed to compensate for tumor repopulation over the entire course of treatment, which is linear in the total treatment time. The constraint (\ref{EqDoseCons}) ensures that the $N$ stage treatment results in the same fraction of surviving tumor cells as a single stage treatment with prescription dose $D^P$. Equations (\ref{a}-\ref{delta}) determine the time evolution of the tumor state as described in section \ref{SecTumorModel}. It should be noted that the initial sparing factor $\delta$ factors out of the objective function. Thus, the optimal dose levels $d_i$ and optimal time gaps $t_i$ are independent of the initial sparing factor $\delta$. In order to make all results independent of the initial sparing factor $\delta$, we define the relative cumulative normal tissue dose as
\begin{equation}
d^L = \frac{\sum_{i=1}^N \delta_i d_i}{\delta D^P} = \frac{\sum_{i=1}^N (v_i)^{2/3} d_i}{D^P}
\label{EqDL}
\end{equation}
which is the cumulative normal tissue dose for an $N$-stage treatment divided by the normal tissue dose for a single stage.

\subsection{Model parameters}
It is evident that the model parameters are uncertain and vary between patients and tumor histologies. In order to obtain insight into the optimal design of multi-stage treatments, we study a nominal set of parameters in sections \ref{Sec:2stages} and \ref{Sec:nstages}. The parameter values are summarized in table \ref{Tab:parameters}. In section \ref{SecSensitivity} we perform a sensitivity analysis to study two-stage treatments over a wide range of parameters.

The set of nominal model parameters are motivated by the liver tumor case presented in section \ref{SecCase}. The prescription dose is set to $D^P=60$ Gy. The tumor shrinkage parameter $\tau_s$ was set to 4 months based on the fact that in our clinical example case the tumor shrank within 4 months to less than 50\% of its original size. The repopulation parameter $\tau_g$ was also set to 4 months based on reports from the literature that both primary liver tumors and metastases exhibit a relatively long tumor doubling time of around 3 months or more \cite{barbara1992,nomura1998,amikura1995}. Note that the tumor doubling time equals $\ln 2 \, \tau_g =  0.7 \,\tau_g $. The effective radiosensitivity parameter $\alpha$ was set to $\alpha = 0.4$~Gy$^{-1}$. This is consistent with estimated radiosensitivity parameters of primary liver tumors \cite{zheng2005impact} and a wide spectrum of other primary tumors \cite{fertil1985intrinsic,williams2008quantitative}. 


\begin{table}[hbt]
\begin{centering}
\begin{tabular}{lll}
\hline \hline
Time constant for tumor shrinkage & $\tau_s$ & 4 months \\
Time constant for repopulation  & $\tau_g$ & 4 months \\
Effective radiosensitivity  & $\alpha$ & 0.4 Gy$^{-1}$ \\
Prescription dose & $D^P$ & 60 Gy \\
\hline \hline
\end{tabular}
\caption{Nominal model parameters used in this paper. }
\label{Tab:parameters}
\end{centering}
\end{table}

\section{Two-stage treatments}
\label{Sec:2stages}
\begin{figure*}[htb]
\centering
\subfigure[relative normal tissue mean dose $d^L(d_1,t_1)$]{
\includegraphics[height=5cm]{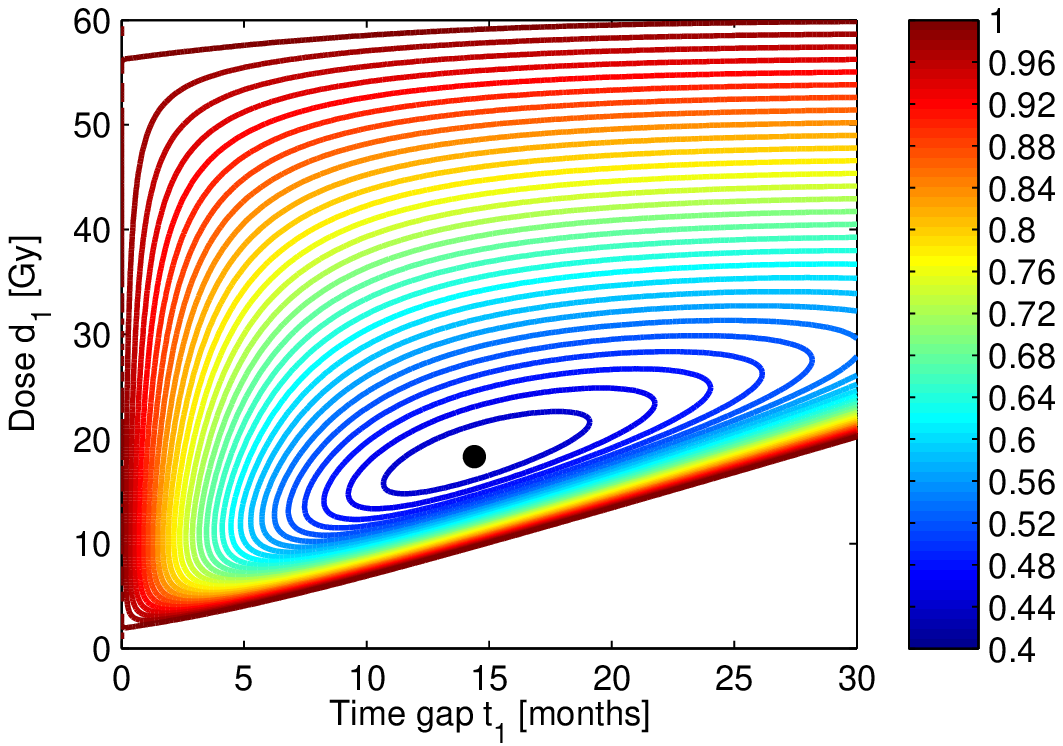}
}
\subfigure[relative tumor volume $v_2(d_1,t_1)$]{
\includegraphics[height=5cm]{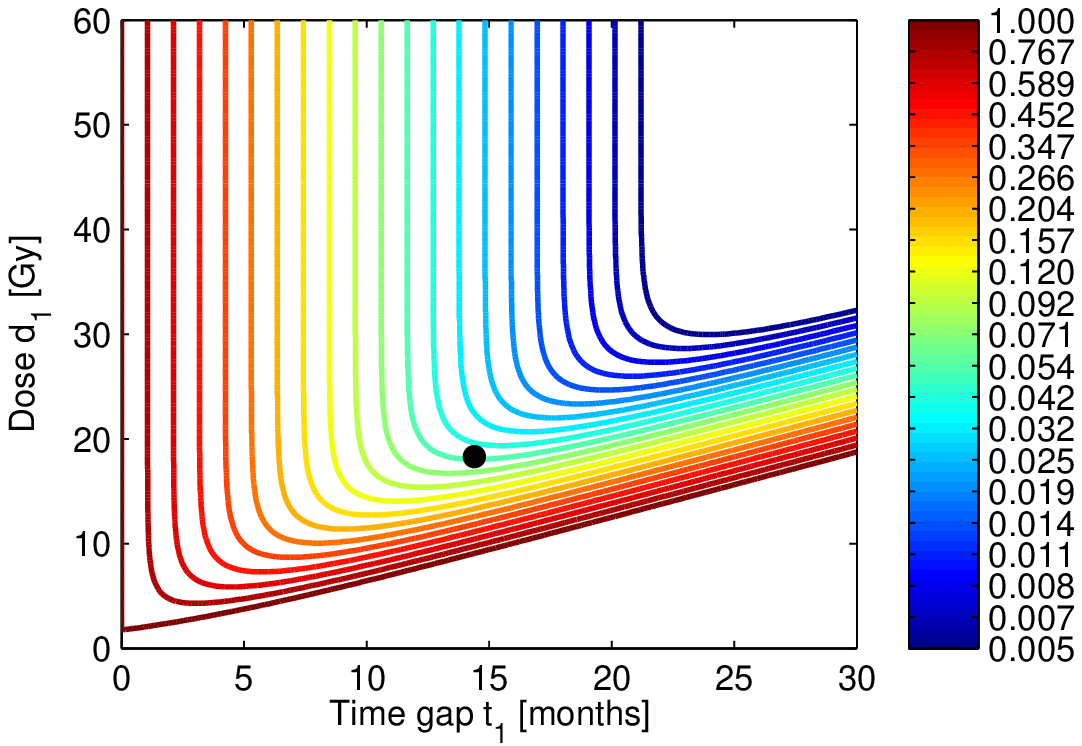}
}
\caption{(a) Cumulative normal tissue mean dose $d^L$ as a function of initial dose $d_1$ and time gap $t_1$ for the parameters in table \ref{Tab:parameters}; (b) Relative tumor volume $v_2 (d_1,t_1)$ in the second stage of the treatment. The black dot marks the location of the optimal solution.}
\label{Fig:2stage}
\end{figure*}

We begin by studying a two-stage treatment by setting $N=2$ in the general model (\ref{EqObj}-\ref{EqInitial}), which may represent the most important application in practice. In this case, the number of decision variables is reduced to two, namely
\begin{itemize}
\item[1.] the dose $d_1$ delivered to the tumor in the first stage, and 
\item[2.] the time gap $t_1$ between the first and second stage.
\end{itemize}
Given that we are considering only two stages, the dose in the second stage, $d_2$, is determined by the constraint (\ref{EqDoseCons}) and given by
\begin{equation}
d_2 = \frac{t_1}{\alpha\tau_g} + D^P - d_1 \label{Eq:Dose2nd} 
\end{equation}
We want to determine the time gap $t_1$ and the doses $d_1$ and $d_2$ such that the cumulative normal tissue dose after the end of the second stage is minimized. Since we have only two free decision variables, $d_1$ and $t_1$, we can simply plot the cumulative normal tissue dose as a function of these. The result is shown in figure \ref{Fig:2stage}a for the set of parameters in table~\ref{Tab:parameters}. It is seen in figure \ref{Fig:2stage}a that the cumulative normal tissue dose has a unique minimum corresponding to the following treatment regimen:
\begin{itemize}
\item[] Dose levels: $d_1=18.3$ Gy, $d_2=50.7$ Gy 
\item[] Time gap: $t_1 = 14.4$ months 
\item[] Relative cumulative normal tissue dose: $d^L = 0.42$. 
\end{itemize}
In our model, it is optimal to deliver a relatively small initial dose of 18.3 Gy. After a time gap of 14.4 months a larger dose of 50.7 Gy is delivered. The cumulative normal tissue dose for this treatment regimen is 42\% of the dose in a single stage treatment, thus representing a substantial normal tissue dose reduction.

\subsection{Qualitative analysis}
\label{Sec:Qualitative}
We can further qualitatively interpret figure \ref{Fig:2stage}a in order to discuss the different competing factors that determine the optimal treatment regimen. For that purpose, figure \ref{Fig:2stage}b shows the relative tumor volume $v_2 (d_1,t_1)$ in the second stage. 

\paragraph{Limit of long time gaps and small initial doses:} It is intuitive that a treatment is suboptimal if the tumor volume in the second stage is larger than in the first stage. This is the case for low doses $d_1$ and long time gaps $t_1$ where the tumor volume is dominated by repopulation. In figure \ref{Fig:2stage}b we plot the line where $v_2(d_1,t_1) = 1$ (corresponding to $\delta_2(d_1,t_1) = \delta$). This line (lowermost dark red line) limits the range of reasonable treatment regimens away from large time gaps and low initial doses. The line in $(d_1,t_1)$ space where $v_2(d_1,t_1)=1$ is approximately a straight line with slope $1/(\alpha\tau_g)$, which corresponds to the extra dose added for repopulation per unit time interval.


\paragraph{Optimal initial dose:} For any given time gap $t_1$, increasing the initial dose $d_1$ decreases the surviving fractions of tumor cells. However, for large $d_1$, further increases do not translate into noticeable reduction in the tumor volume, which becomes dominated by the removal of dead cells. This is seen as the vertical isolines in figure \ref{Fig:2stage}b. On the other hand, a large initial dose means that the dose in the second stage is small. Thus, the ability to take advantage of a lower sparing factor is reduced, and as seen in figure \ref{Fig:2stage}a, the mean normal tissue dose increases (approximately linearly) for large initial doses. Thus, the initial dose $d_1$ has to be large enough to provide tumor shrinkage, but small enough to capitalize on the tumor shrinkage in the second stage. For our model, the optimal trade-off is achieved if approximately 30\% of the dose is delivered in the first stage.

\paragraph{Optimal time gap:} For small time gaps, the treatment regimen is suboptimal because the tumor is still in its shrinking phase. The dose reduction in the second stage can be improved by waiting longer for the tumor to shrink further. For small time gaps, this advantage compensates for the adverse effect of the extra dose added to compensate for repopulation. Intuitively, we expect the time gap should be chosen such that the second stage is delivered when the tumor size reaches the minimum. It is straightforward to see that this is strictly true if no extra dose is added for repopulation in (\ref{Eq:Dose2nd}). Otherwise, the second stage should be delivered when the tumor is still in its shrinking phase as shown in appendix \ref{Sec:App_opt_gap}. The numerical results indicate (figure \ref{Fig:2stage}b) that the optimal time to deliver the second stage is approximately when the tumor size reaches the minimum.

\section{Sensitivity analysis} 
\label{SecSensitivity}
The predicted benefit of a two-stage treatment over a single-stage treatment depends on model assumptions and parameters. In this section, we study the benefit of two-stage treatments over a wide range of parameter values and consider the dependence of the initial dose $d_1$ and the time gap $t_1$ on parameter values. 



\subsection{Repopulation and shrinkage} 
\begin{figure*}[htb]
\centering
\subfigure[fast repopulation]{
\includegraphics[height=5cm]{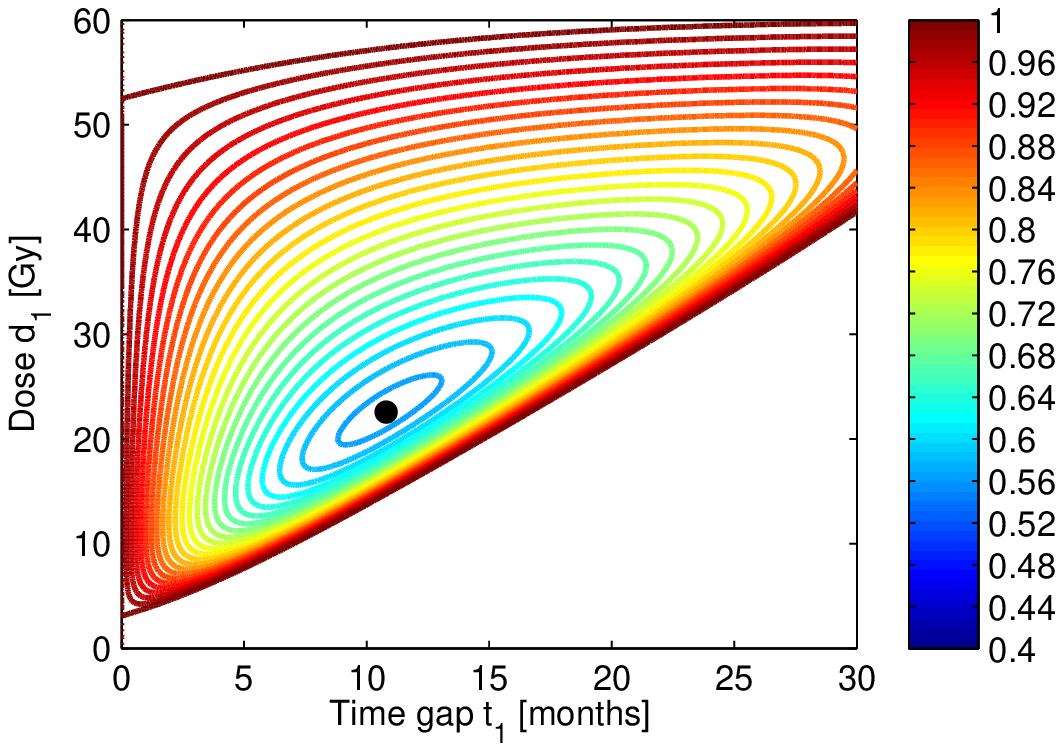}
}
\subfigure[unfavorable parameters]{
\includegraphics[height=5cm]{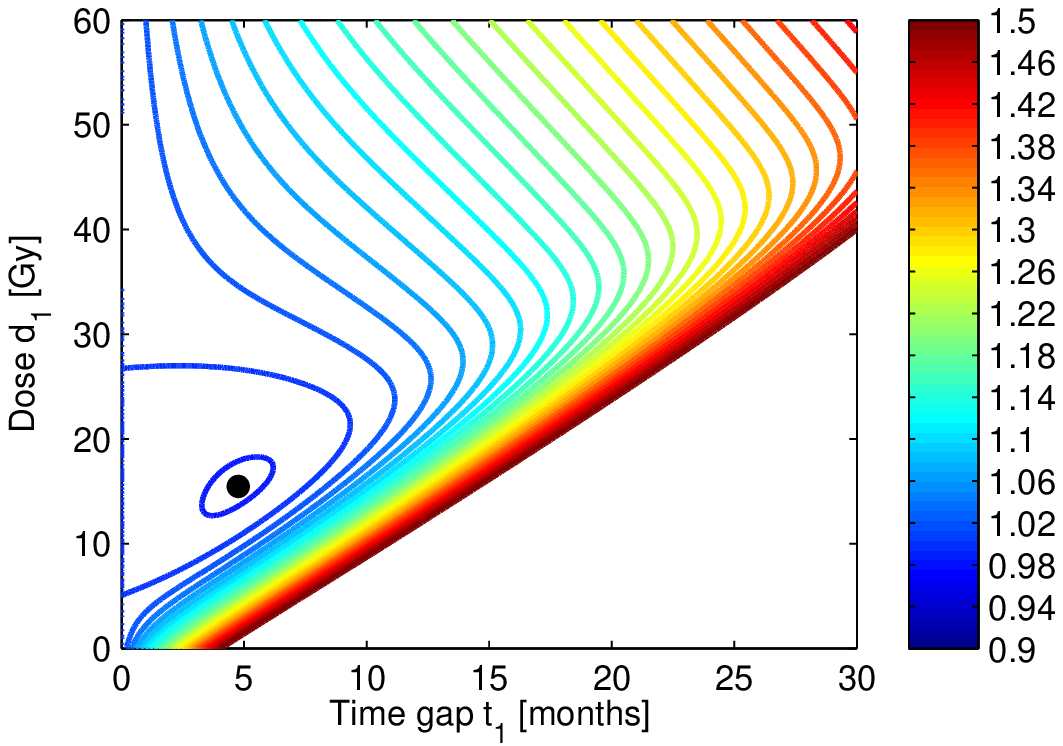}
}
\caption{Cumulative normal tissue mean dose $d^L$ as a function of initial dose $d_1$ and time gap $t_1$: (a) for $\tau_g=2$ months; (b) for $\tau_g=2$ months, $V_{min}=0.1$, and $q=0.16$. The remaining parameters are chosen according to table \ref{Tab:parameters}.}
\label{FigUnfavorable}
\end{figure*}

Fast tumor regression (small $\tau_s$) is in favor of two-stage treatments. Thus, decreasing $\tau_s$ would further increase the projected benefit of two-stage treatments, and suggest shorter treatment breaks $t_1$. Likewise, slow repopulation (large $\tau_g$) favors two-stage treatments.  On the other hand, fast tumor cell repopulation (small $\tau_g$) works against the benefit of two-stage treatments. Decreasing $\tau_g$ leads to shorter time gaps $t_1$, and higher cumulative normal tissue dose $d^L$. Figure \ref{FigUnfavorable}a shows the cumulative normal tissue dose as a function of $d_1$ and $t_1$ for a repopulation time constant of $\tau_g = 2$ months (twice the value compared to figure \ref{Fig:2stage}). As a consequence, the slope of the line $v_2(d_1,t_1) = 1$, which limits beneficial treatment regimens away from large $t_1$ and small $d_1$, is doubled. This pushes the optimal treatment to a shorter time gap $t_1=10.8$ months and a slightly higher initial dose $d_1=20.9$ Gy, and increases the achievable cumulative normal tissue dose to $d^L=0.55$ (compared to $t_1=14.4$, $d_1=18.3$, and $d^L=0.42$ for the nominal parameters).\\

\begin{figure*}[htb]
\centering
\includegraphics[height=7cm]{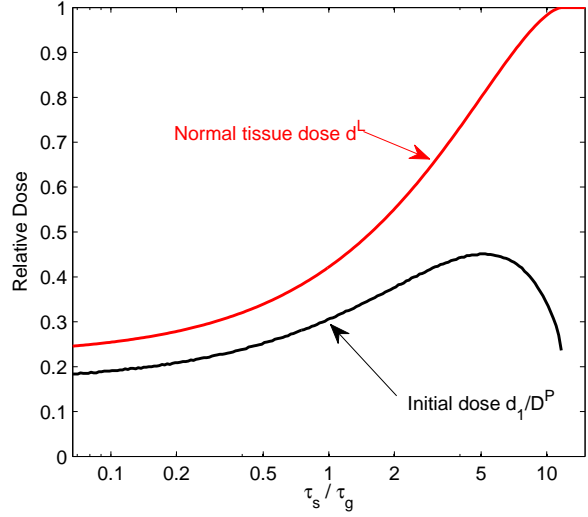}
\caption{Dependence of the cumulative normal tissue dose $d^L$ (red line) and the optimal initial dose $d_1/D^P$ (black line) on the ratio of shrinkage and repopulation time constant. The remaining parameters are chosen as in table \ref{Tab:parameters}.}
\label{FigSensitivityTauratio}
\end{figure*}

In the tumor model, the treatment break $t_1$ only appears in ratios $t_1/\tau_s$ and $t_1/\tau_g$. Thus, if time is measured in units of the repopulation constant $\tau_g$, it becomes apparent that the results for $d_1$ and $d^L$ are independent of the absolute values of $\tau_s$ and $\tau_g$, and only depend on the ratio of shrinkage and repopulation time constant $\tau_s/\tau_g$ \footnote{This also means that, for any fixed ratio $\tau_s/\tau_g$, the treatment break $t_1$ increases linearly with the absolute values of the time constants, but the proportionality constant depends on $\tau_s/\tau_g$.}. Figure \ref{FigSensitivityTauratio} shows the optimal dose delivered in stage one and the cumulative normal tissue dose as a function of the ratio $\tau_s/\tau_g$. There is a wide range of parameter values for which there is a benefit for two-stage treatments. Only for very fast repopulation, if $\tau_g$ is approximately 11 times smaller than $\tau_s$, the optimal treatment break becomes zero and a single stage treatment becomes optimal. When repopulation is approximately 5 times faster than shrinkage, the dose delivered in stage one has a maximum at 45\% of the prescription dose. When $\tau_s$ and $\tau_g$ are equal, approximately 30\% of the dose is delivered in stage one. For very fast shrinkage, approximately 20\% of the dose is delivered in the first stage.


\subsection{Radiosensitivity}
\begin{figure*}[htb]
\centering
\includegraphics[height=7cm]{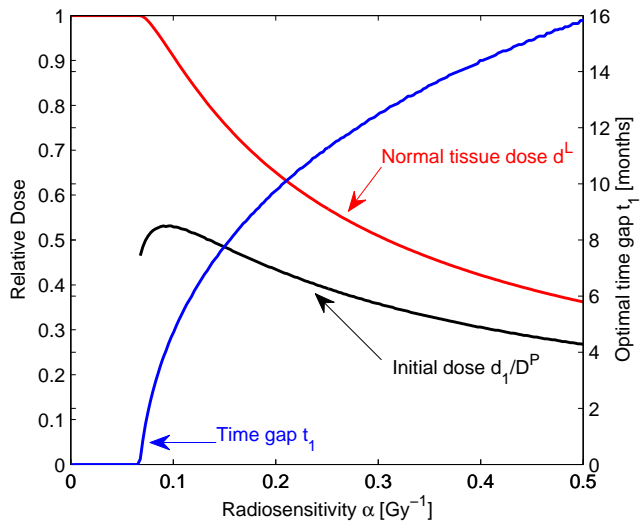}
\caption{Dependence of the cumulative normal tissue dose $d^L$ (red line), the optimal time gap $t_1$ (blue line), and the optimal initial dose $d_1/D^P$ (black line) on the radiosensitivity parameter $\alpha$. The remaining parameters are chosen as in table \ref{Tab:parameters}.}
\label{FigSensitivityAlpha}
\end{figure*}

Figure \ref{FigSensitivityAlpha} shows the dependence of the normal tissue dose and the decision variables on the radiosensitivity parameter $\alpha$. The benefit of a two stage treatment decreases for radioresistent tumors, which is expected because a given initial dose $d_1$ leads to less tumor shrinkage that can be exploited in the second stage. However, only for radiosensitivity parameters of $\alpha \approx 0.07$ Gy$^{-1}$ (corresponding to a surviving fraction of 0.87 at 2 Gy) a single stage becomes optimal. In addition, the optimal initial dose $d_1$ depends only mildly on the radiosensitivity parameter. 


\subsection{Normal tissue sparing factor model} 
\label{SecSensitivityQ}
\begin{figure*}[htb]
\centering
\includegraphics[height=6cm]{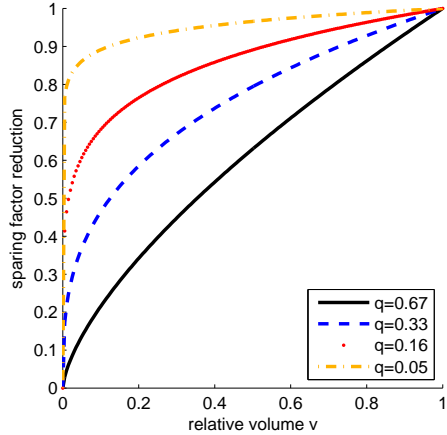}
\caption{Sparing factor reduction $v^q$ as a function of volume reduction $v$ for different exponents $q$. A value of $q=0.05$ corresponds approximately to the sparing factor model at which a single stage treatment becomes optimal for the parameters in table \ref{Tab:parameters}.}
\label{FigSparingModel}
\end{figure*}

In the main text of this paper, we used the generic model in equation  \ref{Eq:delta} for the relation of the sparing factor $\delta_i$ and reduction in tumor volume $v_i$. This model does not account for the patient specific geometry of tumor and dose-limiting normal tissue, nor does it take into account the beam arrangement. For a patient at hand, the relation between $\delta_i$ and $v_i$ may be different. In order to assess the benefit of two-stage treatments for a wider range of sparing factor models, we consider the generalized model
\begin{equation}
\delta_i = \delta \cdot v_i^q
\end{equation}
where $q=2/3$ corresponds to the model used throughout the paper. For smaller values of the exponent, the reduction in the sparing factor as a function of tumor volume becomes more shallow around the initial sparing factor $\delta$ (see figure \ref{FigSparingModel}). Thus, values $q < 2/3$ correspond to situations that are less favorable for two-stage treatments. For $q \to 0$ there is no reduction in the sparing factor until the tumor volume becomes zero. \\

\begin{figure*}[htb]
\centering
\includegraphics[height=7cm]{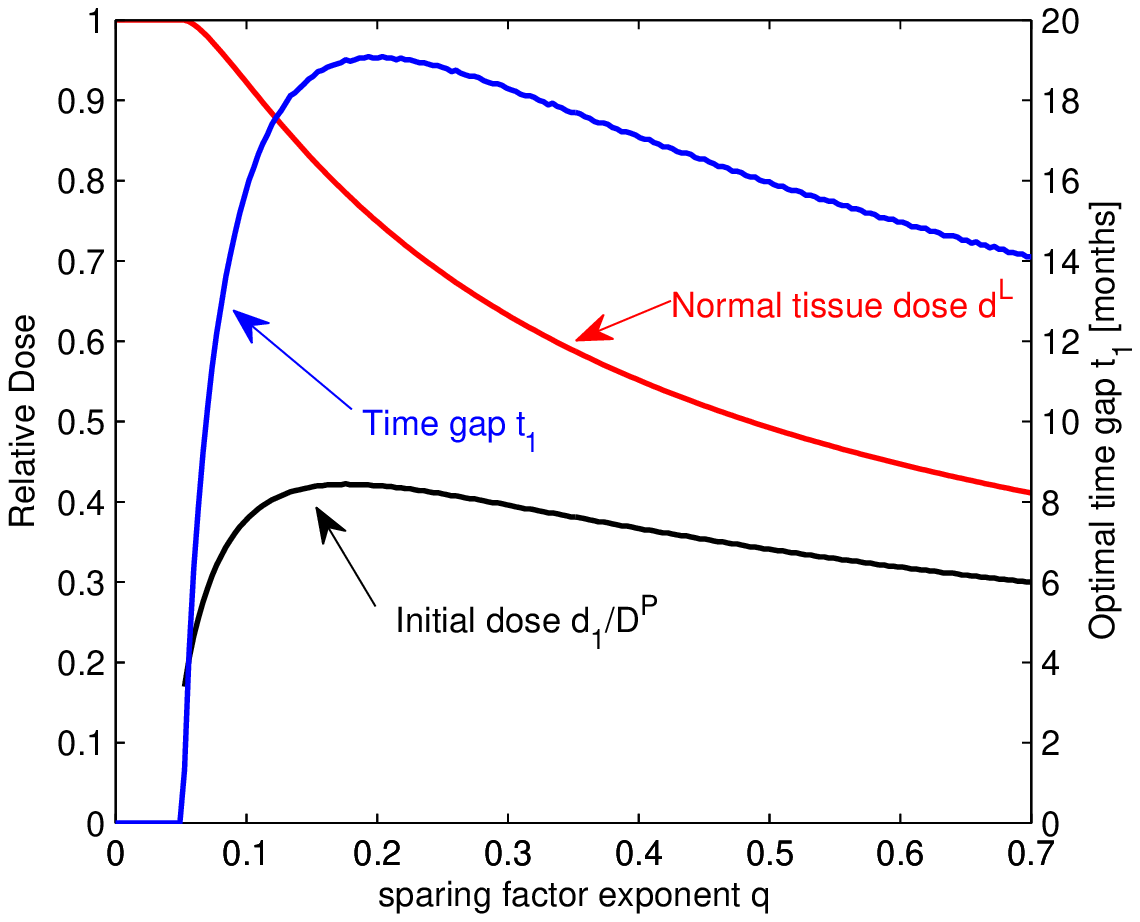}
\caption{Dependence of the cumulative normal tissue dose $d^L$ (red line), the optimal time gap $t_1$ (blue line), and the optimal initial dose $d_1/D^P$ (black line) on the sparing factor model parameter $q$. The remaining parameters are chosen as in table \ref{Tab:parameters}.}
\label{FigSensitivityExponent}
\end{figure*}

Figure \ref{FigSensitivityExponent} shows the decision variables $t_1$ and $d_1$ and the cumulative normal tissue dose as a function of the exponent $q$. As expected, the benefit of two-stage treatments decreases for smaller exponents because a tumor volume reduction translates into only a minor sparing factor reduction. However, two-stage treatments remain optimal up to exponents of $q\approx0.05$. In addition, the dose delivered in the first stage depends only mildly on the sparing factor model.

\subsection{Minimum target volume} The above results assume that the target volume in the second stage is determined by the number of viable and dead (but not yet removed) tumor cells. In reality, the residual target volume may be larger if the location of the remaining viable cells is uncertain. To address this concern, we considered a modified version of equation (\ref{Eq:delta}) in which the sparing factor is calculated according to 
\begin{equation}
\delta_2(d_1,t_1) = \delta \left[ v_2(d_1,t_1) + \left(1-v_2(d_1,t_1) \right) V_{min}\right]^{2/3}
\label{Eq:delta_vmin}
\end{equation}
where $V_{min} \in [0,1]$ is the relative minimum target volume that is to be treated in the second stage. $V_{min}$ can be thought of as a clinical target volume (CTV) containing areas of potential microscopic disease beyond a visible mass of necrosis and residual tumor. \\

\begin{figure*}[htb]
\centering
\includegraphics[height=7cm]{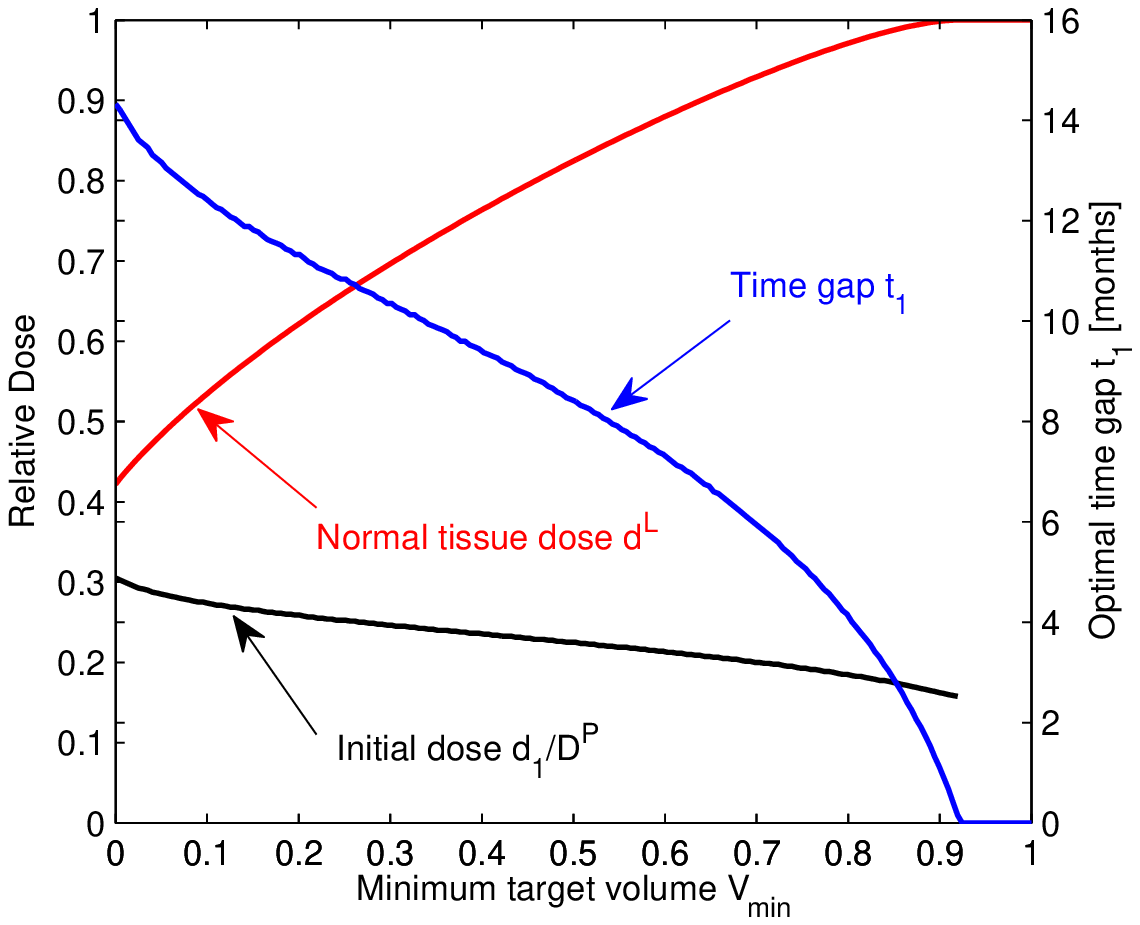}
\caption{Dependence of the cumulative normal tissue dose $d^L$ (red line), the optimal time gap $t_1$ (blue line), and the optimal initial dose $d_1/D^P$ (black line) on the minimum target volume $V_{min}$. The remaining parameters are chosen as in table \ref{Tab:parameters}.}
\label{FigSensitivityVmin}
\end{figure*}

Figure \ref{FigSensitivityVmin} shows the cumulative normal tissue dose and the decision variables as a function of the relative minimum target volume $V_{min}$. As expected, the advantage of a two-stage treatment in terms of normal tissue sparing decreases with increasing $V_{min}$. Furthermore, the optimal time gap decreases. A minimum target volume results in a minimum sparing independent of the time gap and the initial dose. Hence, prolonging the treatment through large time gaps $t_1$ approximately leads to a linear increase $t_1 \delta V_{min}^{2/3} /\left(\alpha \tau_g\right)$ of the mean normal tissue dose. Nevertheless, figure \ref{FigSensitivityVmin} shows a substantial projected benefit of a two-stage treatment for large residual target volumes. For example, a fairly large residual target volume of 20\% yields a relative normal tissue dose of $d^L=0.61$. Only in the case that 90\% of the initial target volume is not subject to shrinkage, a single stage treatment becomes optimal (figure \ref{FigSensitivityVmin}). The optimal initial dose $d_1$ depends only mildly on the minimum target volume. \\

Clearly, single stage treatments become optimal if all parameters are chosen as to work against the benefit of two-stage treatments, i.e. fast repopulation, large residual treatment volume, small sparing factor reduction. This is illustrated in figure \ref{FigUnfavorable}b for the parameters $\tau_g=2$ months, $V_{min}=0.1$, and $q=0.16$. In this case, a two-stage treatment is still optimal ($t_1=4.76$ months, $d_1=15.5$ Gy), but the benefit is negligible ($d^L=0.98$). Further worsening one of the parameters eliminates the local minimum and makes single stage treatments more favorable.

\section{Multi-stage treatments}
\label{Sec:nstages}
In this section we discuss treatment schedules consisting of more than two stages. In section \ref{Sec:2+1stages}, we start by discussing the benefit of adding a third treatment stage half way into the optimal two-stage treatment. In section \ref{Sec:opt3stages}, we consider the optimal three-stage treatment for optimal time gaps. In section \ref{Sec:Qstages} we obtain insight into quasi-continuous treatments in which the patient is irradiated in many stages at regular time intervals.

\subsection{Adding a third stage at fixed time intervals}
\label{Sec:2+1stages}
\begin{figure*}[htb]
\centering
\includegraphics[height=5cm]{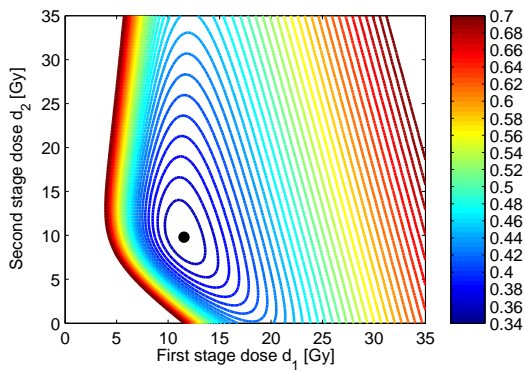}
\caption{Cumulative normal tissue mean dose for a 3-stage treatment as a function of the dose levels $d_1$ and $d_2$ for the parameters in table \ref{Tab:parameters} and fixed time gaps of 7.2 months.}
\label{Fig:3stages}
\end{figure*}

We consider the set of nominal parameters and fixed time gaps $t_1=t_2=7.2$ months, such that the total treatment time $t_1+t_2$ remains equal to the optimal two stage treatment (14.4 months). The optimal dose levels for the three treatment stages are as follows:
\begin{itemize}
\item[] Dose levels: $d_1=11.6$ Gy, $d_2=9.8$ Gy, $d_3=47.6$ Gy 
\item[] Time gaps: $t_1 = t_2 = 7.2$ months 
\item[] Relative cumulative normal tissue dose: $d^L=0.35$  
\end{itemize}
It is observed that adding a third stage half way into the optimal two-stage treatment yields a moderate improvement. The relative cumulative normal tissue dose decreases from 42\% to 35\%. Figure \ref{Fig:3stages} shows the cumulative normal tissue dose as a function of the dose levels in stage 1 and 2. In comparison to the optimal two-stage treatment, we note that the dose delivered in the last treatment stage remains approximately the same. Instead, the dose that is delivered initially in the two-stage treatment is partly shifted into the intermediate stage of the three-stage treatment. For a fixed dose in the final stage, the distribution of dose over stage 1 and 2 is driven by two competing objectives: (1) delivering a large dose in stage 1 minimizes the tumor volume in the final stage, and (2) any dose delivered in the second stage can take advantage of the tumor shrinkage that occurs over the first time gap. This leads to the compromise found in the solution above. 

\subsection{The optimal three-stage treatment}
\label{Sec:opt3stages}
We now consider the optimal three-stage treatment by determining the optimal doses $d_1$ and $d_2$ as well as the optimal time gaps $t_1$ and $t_2$. The optimal treatment regimen was obtained by exhaustive search over the four parameters. The following treatment regimen was obtained:
\begin{itemize}
\item[] Optimal dose levels: $d_1=11.9$ Gy, $d_2=15.4$ Gy, $d_3=46.7$ Gy 
\item[] Time gaps: $t_1=8.7$ months, $t_2=13.7$ months 
\item[] Relative cumulative normal tissue dose: $d^L=0.32$  
\end{itemize}
It is observed that optimizing the time gaps of a 3-stage treatment further prolongs the treatment. However, the projected benefit compared to the regimen in section \ref{Sec:2+1stages} is small. A summary of the results above for single stage, two-stage and three-stage treatments is provided in table \ref{Tab:summary}. 
\begin{table}[hbt]
\begin{centering}
\begin{tabular}{lrrrrrr}
\hline \hline
 & $d_1$(Gy) & $t_1$(months) & $d_2$(Gy) &  $t_2$(months) & $d_3$(Gy) &  $d^L$ \\
\hline
Single stage & 60.0 & -- & --  & --  & --  &  1.00 \\
2-stage  & 18.3 & 14.4  &  50.7  &  -- &  -- & 0.42 \\
3-stage $(t_1=t_2)$ & 11.6 & 7.2 & 9.8 & 7.2 & 47.6 & 0.35 \\
3-stage (optimal)   & 11.9 & 8.7  & 15.4   & 13.7   &   46.7   &  0.32   \\
\hline \hline
\end{tabular}
\caption{Comparison of optimal dose levels and time gaps for two- and three-stage treatments in comparison with a single stage treatment. The reduction of dose $d^L$ to healthy tissue is shown in the last column. }
\label{Tab:summary}
\end{centering}
\end{table}

\subsection{Quasi-continuous treatments}
\label{Sec:Qstages}
\begin{figure*}[htb]
\centering
\includegraphics[height=5cm]{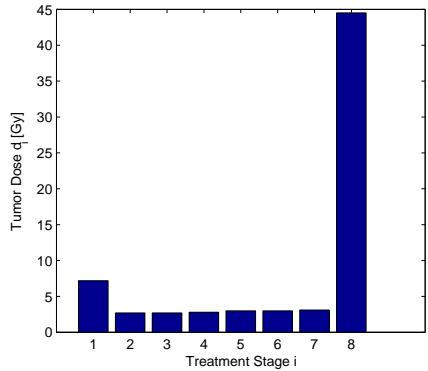}
\caption{Dose levels for the optimal 8-stage treatment delivered over 14.4 months (i.e. the time gap between two consecutive stages is slightly over 2 months).}
\label{Fig:8stages}
\end{figure*}

Finally we now consider a treatment regimen in which the patient is treated quasi continuously, i.e. at relatively short regular intervals for a fixed maximum treatment period. We consider fixed time gaps $t_i=t$ for all of $N$ treatment stages such that the dose levels $d_i$ are the only decision variables. To optimize the dose levels, we use dynamic programming as described in appendix \ref{SecDP}.  To illustrate the main findings regarding quasi-continuous treatment regimens, we consider an 8 stage treatment with fixed time gaps of slightly over 2 months, such that the total treatment time is 14.4 months (corresponding to the time gap of the optimal two-stage treatment). The optimal dose levels are shown in figure \ref{Fig:8stages}. The treatment starts with a dose of 7 Gy in the first stage. Between stage 2 and 7, a relatively low maintenance dose of approximately 3 Gy is delivered. Most of the dose (44 Gy) is delivered in the last stage when the tumor size is smallest. The treatment regimen shown in figure \ref{Fig:8stages} yields a cumulative normal tissue dose of 0.33\%. This is slightly lower than the three-stage treatment in section \ref{Sec:2+1stages} for the same overall treatment duration. Similar treatment schedules (i.e. a sizeable initial dose, followed by a low maintenance dose, and a high final boost dose) have been observed over a wide range of model parameters.

\section{Discussion}
\label{SecDisc}

Multi-stage radiotherapy has long been considered as a potential treatment regimen. Motivations for multi-stage radiotherapy include: First, radiobiological reasons, e.g. the regeneration of healthy tissues over the treatment break; and second, physical reasons, e.g. better normal tissue sparing due to a shrinking field size. Previous clinical trials on multi-stage radiotherapy were mostly motivated by radiobiological reasons. Since these trials failed to establish a benefit, multi-stage radiotherapy is rarely applied in current clinical practice. However, advances in imaging as well as advanced treatment techniques such as IMRT and SBRT, suggest taking a fresh look at multi-stage treatments for selected sites due to physical reasons. 


\subsection{Main findings of the paper} 

\paragraph{Two-stage treatments:} We analyzed a model of a tumor which incorporates three factors: radiation-induced cell kill, tumor shrinkage, and tumor repopulation. Our analysis has focused on two-stage treatments. The model allows us to estimate the potential normal tissue dose reduction achievable by treating a smaller tumor in the second stage. In addition, the optimal initial dose that should be delivered in the first treatment stage and the optimal time gap between first and second stage are obtained. It is shown that there is potential for normal tissue dose reduction for a wide range of model parameters. For favorable parameters, the projected benefit of two-stage treatments is substantial and exceeds a normal tissue dose reduction by more than 50\%. The model parameters will be uncertain and dependent on the treatment site. This applies to both the radiobiological parameters for radiosensitivity, repopulation and shrinkage, as well as the parameters to relate tumor shrinkage to normal tissue sparing. A sensitivity analysis with respect to the parameters revealed that the initial dose delivered in the first treatment stage is approximately 30\% to 45\% of the prescribed tumor dose that would be delivered in a single-stage treatment. The second stage should be delivered just before the tumor size reaches a minimum.


\paragraph{Insight into the optimal quasi-continuous treatment: } We obtained insight into the optimal quasi-continuous treatment in which the patient is treated at short regular time intervals for a fixed total treatment time. Qualitatively, the model suggests a treatment policy in which the treatment is initiated with a moderate initial dose, followed by an approximately constant low maintenance dose, and a final dose boost in the last stage. The projected benefit of adding additional treatment stages is relatively small compared to the benefit of two stages over a single stage. The small projected benefit of additional stages, together with potential concerns regarding the validity of the tumor response model for very small dose levels in each stage, suggest that a two-stage approach is a reasonable consideration for clinical implementation.

\subsection{Potential applications for multi-stage radiotherapy}
\label{SecApplication}
Our work in this paper is primarily motivated by patients with large liver tumors as illustrated in section \ref{SecCase}. However, with continuing improvements in image guidance and dose delivery precision, the potential of multi-stage radiotherapy (motivated by tumor shrinkage) could be reevaluated. This analysis can be guided by the following criteria that favor multi-stage treatments:
\begin{itemize}
\item[1.] The disease site exhibits significant tumor regression, which translates into a smaller treatment volume in the second phase. This suggests treatment sites where a large portion of the target volume represents gross disease that is subject to regression. In contrast, treatment sites where most of the target volume represents potential microscopic disease (where no reduction in the target volume can be exploited) are less promising candidates for multi-stage treatments. 
\item[2.] The adverse effects of protracting the treatment are small. This suggests tumors that have relatively slow tumor repopulation rates, such as liver cancers \cite{barbara1992}, as potential candidates. In contrast, tumors exhibiting fast repopulation (or accelerated repopulation) are less likely to be candidates for multi-stage therapy. In addition, for some tumor entities, potential metastatic spread of disease may represent an additional concern for prolonging treatment.
\end{itemize}
It is evident that for any treatment site that is a potential candidate for multi-stage radiotherapy based on tumor shrinkage, there will be other clinical factors which play a role in the design of the treatment scheme. In particular, the potential for normal tissue dose reduction due to tumor shrinkage has to be weighed against the possible risks of prolonging the treatment. This would have to be assessed in a disease site specific manner, which is outside the scope of this paper. Thus, this paper demonstrates the potential for normal tissue dose reduction by exploiting tumor shrinkage in multi-stage treatments, suggesting that this should be considered as a factor in designing treatment schemes. However, to devise a concrete application, a broader view involving all clinical factors has to be taken.

\subsection{Application to liver tumors} 
\label{SecLiver}
Since this work was motivated by the management of large liver tumors we provide a more detailed discussion regarding this potential application. For large liver tumors, multi-stage radiotherapy is motivated by several factors:
\begin{itemize}
\item[1.] Dose-volume constraints for the non-involved liver prevent the delivery of an ablative radiation dose in a single course. 
\item[2.] Regrowth of normal liver occurs over the treatment break. As a consequence the total dose that can be delivered to the liver is increased.
\item[3.] Normal tissue dose can be reduced by treating a smaller tumor in the second stage as discussed in this paper.
\end{itemize}
The development of the liver after the first course of radiation will play an important role in designing split-course treatments. That includes the timescale on which regrowth of normal liver occurs as a function of dose delivered in the first stage, the reirradiation tolerance of the liver as a function of the time gap, as well as the spatial location of functional and non-functional liver. In addition, the combination of radiation with other therapies has to be considered. For example, the question arises whether chemotherapy should be continued over the treatment break.\\

In order to further evaluate the role of normal tissue dose reduction due to tumor shrinkage, the relation between tumor shrinkage $v$ and the mean liver dose reduction $d^L$ has to be studied for realistic patient geometries and beam arrangements. The analysis of follow-up imaging data of previously treated liver patients may provide a way to characterize this relation. Follow-up imaging data acquired approximately at 3 months intervals post treatment provide snapshots of the patient geometry for shrunken tumors. Thus, a treatment planning study involving these different time points may allow us to assess the dependence of $d^L$ on $v$. Likely, this relation will depend on tumor size and location within the liver. \\

For the nominal parameters in table \ref{Tab:parameters}, the model predicts a very long treatment break of 14 months. Whether this is realistic for liver metastasis can be questioned in light of the observed median survival and the time to progression in these patients, which is comparable to this time span \cite{hoyer2012radiotherapy,nair2014}. This may indicate that the nominal model parameters are not adequate for most liver patients. In addition, the model does not account for effects such as varying radiosensitivity of tumor cells or accelerated repopulation. The analysis of follow-up imaging may provide information regarding tumor shrinkage and regrowth over time. However, in practice, two-stage treatments would not depend on an accurate prediction of the treatment gap. Instead, repeated imaging of the patient (e.g. every 1-2 months) can monitor tumor shrinkage over time. If tumor regrowth is observed earlier than expected, the treatment break can be shortened. \\

The model further predicts that the dose delivered in the first stage should be in the order of 30\% to 45\% of the prescription dose. For large liver tumors, which was the original motivation for this work, the interpretation of this result should take into account that the desired prescription dose is higher than what can be safely delivered in a single stage treatment. For example, if the cumulative prescription dose to a large liver tumor is 60 Gy, it may be that only 30 Gy can be delivered in the first stage to max out dose-volume constraints for the liver. In this case, the model suggests a moderate dose reduction to 20 to 25 Gy in the first stage. Thus, the dose reduction in the first stage is less than it may appear. \\

\section{Conclusion}
The treatment of large liver tumors with radiation faces the problem that dose-volume constraints for the non-involved liver prevent the application of a curative radiation dose in a single stage. This motivates the exploration of alternative treatment regimens in which the patient is irradiated in multiple stages seperated by months. In this paper, we assess the potential benefit of exploiting tumor shrinkage that occurs during the treatment break. The reduced target volume in the second stage allows for smaller radiation fields, which translates into a dose reduction in the healthy tissue. We are concerned with temporal optimization of multi-stage radiotherapy. By studying a dynamic model of tumor growth and response to therapy, we obtain insight into the optimal treatment schedule, i.e. the optimal doses delivered in each stage and the treatment breaks in between stages. For a two-stage treatment, the second stage should be delivered just before the tumor size reaches its minimum and repopulation overcompensates shrinkage. The model suggests that the dose delivered in the first stage should approximately be one third of the prescribed dose. Furthermore, we applied dynamic programming in order to determine the optimal treatment schedule in a quasi-continuous treatment where the patient is treated many times at regular time intervals. In this case, the optimal treatment policy consists of a moderate initial dose, followed by a low maintenance dose, and a large boost dose in the last stage. The true benefit of exploiting tumor shrinkage in multi-stage treatments will have to be assessed using patient data while taking the patient and beam geometry into account. However, the large reduction in normal tissue dose by more than a factor of two observed in the model suggests that tumor shrinkage may be an important factor in designing multi-stage treatments.\\

\appendix

\section{On the optimal time gap}
\label{Sec:App_opt_gap}
We consider a two-stage treatment and proof that the second stage should be delivered just before tumor size reaches a minimum. To that end, we calculate the partial derivative of the normal tissue dose $d^L(d_1,t_1)$ in equation (\ref{EqDL}) with respect to the time gap $t_1$.  Using equation (\ref{Eq:Dose2nd}) for the dose in the second stage, and noting that the dose in the first stage is independent of $t_1$, we have 
\begin{equation}
\frac{\partial d^L}{\partial t_1} = \frac{1}{\delta D^P} \left[ \frac{\partial \delta_2(d_1,t_1)}{\partial t_1} d_2(d_1,t_1) + \frac{\delta_2(d_1,t_1)}{\alpha \tau_g} \right]
\end{equation}
At the optimal time gap we have $\partial d^L/ \partial t_1 = 0$. Since $\delta_2(d_1,t_1)>0$ it follows that $\partial \delta_2(d_1,t_1) / \partial t_1$ has to be negative in order to fulfill that condition. This means that the second stage is to be delivered while the tumor is still in the shrinking phase. Intuitively this makes sense since there is a price for delaying the treatment due to the repopulation dose correction. However, as seen in figure \ref{Fig:2stage}b this effect is small; the second stage is delivered when the tumor size is very close to its minimum.

\section{Dynamic programming formulation}
\label{SecDP}
In this section, we provide a dynamic programming (DP) formulation used to obtain the optimal $N$-stage treatment regimen for fixed time gaps $t_i=t$. For a general introduction to DP we refer the reader to \cite{bertsekas2005}. We make the associations below:
\begin{itemize}
\item[] {\bf Control variables:} The control variable at each stage is the dose $d_i$ delivered to the tumor.
\item[] {\bf Cost function:} The cost incurred at stage $i$ is given by the dose $\delta_i d_i$ delivered to the healthy tissue. The goal is minimize the cumulative normal tissue dose.
\item[] {\bf State variables:} At every stage, the state of the tumor is fully described by the number of viable tumor cells $a_i$ and the number of dead tumor cells $z_i$. The corresponding state dynamics is given in equations (\ref{EqState_a}) and (\ref{EqState_z}). We can further note that the number of living cells is uniquely determined by the cumulative dose delivered to the tumor until stage $i$ via the equation 
\begin{equation}
a_{i} = \exp \left( -\alpha \sum_{j=1}^{i-1} d_j \right) \exp \left(\frac{(i-1) t}{\tau_g} \right) 
\end{equation}
and does not depend on the history of dose delivery. Thus, to characterize the state, it is sufficient to keep track of the cumulative tumor dose and the number of dead tumor cells.
\end{itemize}
Since we only have two continuous state variables and one continuous control variable, the DP problem can be solved through discretization and by applying the standard DP algorithm. In this work, we use a discretization of 0.1 Gy in the control variables and the cumulative dose state; for the relative number of dead tumor cells we use a discretization of $0.001$.


\bibliographystyle{unsrt}

\end{document}